\documentclass{article}
\usepackage[utf8]{inputenc}
\usepackage[normalem]{ulem}
\usepackage{graphicx} 
\usepackage{xcolor}
\usepackage{tablefootnote}

\usepackage[font=small,labelfont=bf]{caption}
\usepackage[square,numbers]{natbib}
\usepackage{hyperref}
\usepackage{enumitem}
\usepackage{authblk}
\usepackage{amsmath}

\title{Capability enhancement of the X-ray micro-tomography system via ML-assisted approaches}
\author[1]{Dhruvi Shah}
\author[1]{Shruti Mehta}
\author[2,3]{Ashish Agrawal}
\author[3,4]{Shishir Purohit}
\author[1]{Bhaskar Chaudhury \thanks{Corresponding Author : bhaskar\_chaudhury@daiict.ac.in}}
\affil[1]{Group in Computational Science and HPC, DA-IICT, Gandhinagar, India, 382007}
\affil[2]{Technical Physics Division, Bhabha Atomic Research Centre, Mumbai 400085, India}
\affil[3]{Homi Bhabha National Institute, Training School Complex, Anushakti Nagar, Mumbai 400094, India}
\affil[4]{Institute for Plasma Research, Gandhinagar, India, 382428}
\date{ }
\providecommand{\keywords}[1]{\textbf{\textit{Keywords-}} #1}

\begin{document}
\maketitle
\begin{abstract}
{Ring artifacts in X-ray micro-CT images are one of the primary causes of concern in their accurate visual interpretation and quantitative analysis. The geometry of X-ray micro-CT scanners is similar to the medical CT machines, except the sample is rotated with a stationary source and detector. The ring artifacts are caused by a defect or non-linear responses in detector pixels during the MicroCT data acquisition. Artifacts in MicroCT images can often be so severe that the images are no longer useful for further analysis. Therefore, it is essential to comprehend the causes of artifacts and potential solutions to maximize image quality. This article presents a convolution neural network (CNN)-based Deep Learning (DL) model inspired by UNet with a series of encoder and decoder units with skip connections for removal of ring artifacts. The proposed architecture has been evaluated using the Structural Similarity Index Measure (SSIM) and Mean Squared Error (MSE). Additionally, the results are compared with conventional filter-based non-ML techniques and are found to be better than the latter.}
\end{abstract}


\section{INTRODUCTION}
Computed tomography (CT) or computed axial tomography (CAT) is one of the key technologies that has increased the capability of investigating the interior of any given solid sample and is extensively employed in the field of medical imaging \cite{r1}. The X-rays are incident on the given sample from $360^o$s and the X-rays are collected from the other side of the sample by an X-ray detector \cite{r2}. The collected X-ray will have lesser intensity than the incident X-ray beam, due to the attenuation introduced by the sample, often called projections. These projections, as a function of the angle are subjected to tomographic reconstructions which give out the internal topology of the sample, by either classic analytical or iterative reconstructions \cite{r3,r4}. Generally, the CT/CAT gives high-resolution 3D images of the complete sample (large volume sample, like the human body), whereas the MicroCT, gives very high-resolution images for a particular smaller location within the sample \cite{r5,r6}. The CT/CAT/MicroCT employs X-ray sources as the main radiation source, however, nowadays synchrotron-based X-ray beam sources are also employed for MicroCT applications. The synchrotron X-ray MicroCT (SXMCT) has a wide scope of applicability ranging from material sciences to Palaeontology \cite{r7,r8,r9,r10}. However, the capabilities are restricted due to the incomplete projection information. The complete projection information cannot be achieved due to the radiation dose limitations, space and time constraints, object shape, and malfunctioning in the imaging systems. This incomplete character of the measured data gives rise to the Artifacts in the final reconstructed images \cite{r11,r12,r13,r14}. These artifacts studies and possible removal have become a hot research topic among researchers. Numerous forms of artifacts have been reported for the SXMCT images, namely, beam-hardening artifacts, ring artifacts, motion artifacts, scattering, and many more. These are attributed to different reasons which lead to incomplete measured datasets. Every artifact impacts the reconstruction in a very unique way, and the impact is visible in the final reconstructed image. These artifacts limit the applicability of the MicroCT/SXMCT \cite{r14}.

\begin{figure}[!htbp]
  \centering
  \includegraphics[scale=13]{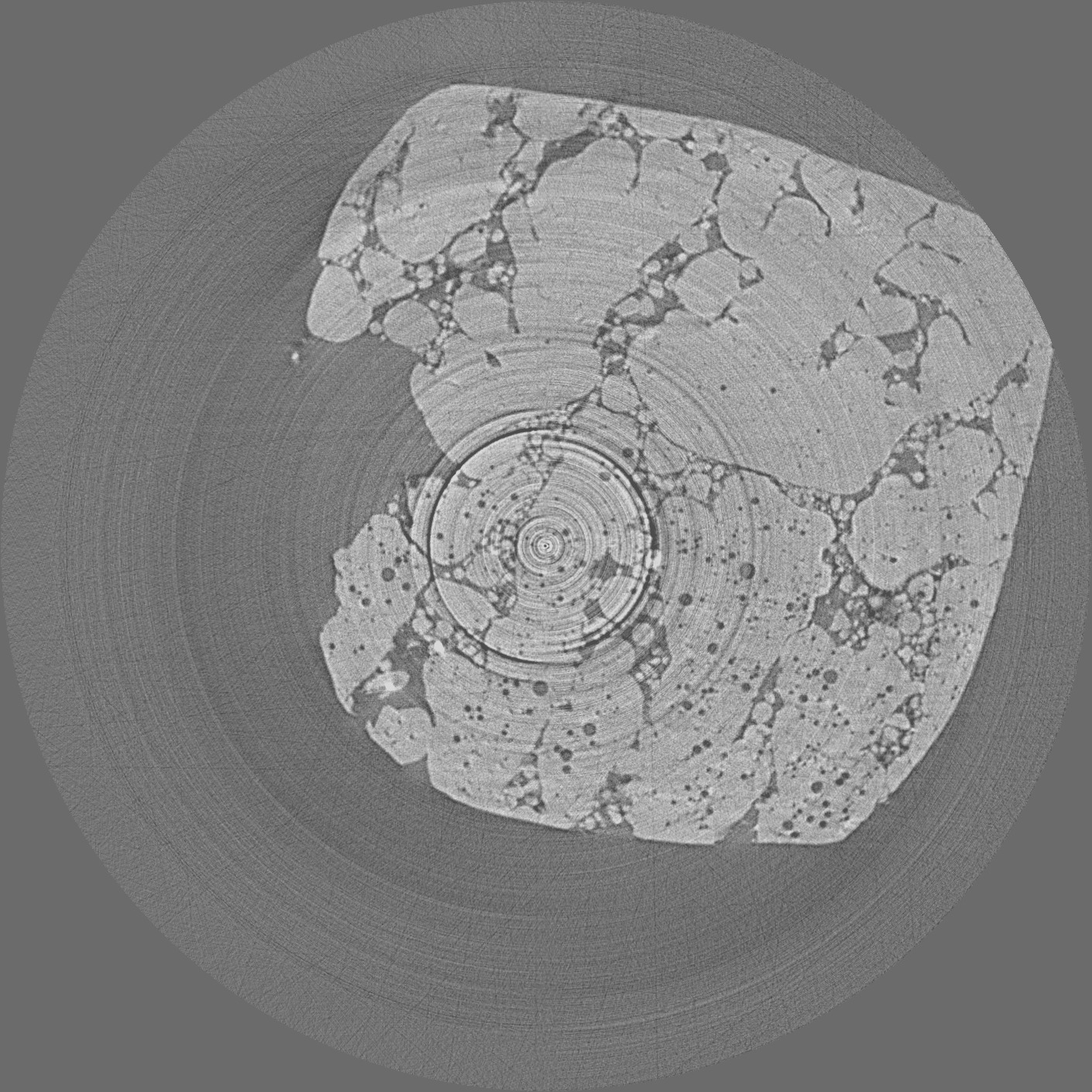}
  \caption{Experimantal micro-ct image containing ring artifacts}
  \label{fig:actual_sample}
\end{figure}

The ring artifacts (RAF) are the most commonly visible and appear to be a ring-like structure of either maximum intensity or the lowest \cite{r15,jnst1}. These rings may have different radii, intensity, and thickness, refer to Figure \ref{fig:actual_sample}. The RAF is the impact of the detector malfunctioning, very specifically, camera pixel malfunction \cite{r16,r17}. When a detector that is supposed to measure the X-ray coming out of the sample is not working properly, it registers a fixed value, which is either the maximum value (hot pixels) or zero in case of a dead pixel. These pixels only give out the most extreme brightness on the scale. The data from such pixels are observed as simple straight lines in the sinogram, and when reconstruction is carried out of these sinograms, RAF are visible as rings in the reconstructed image. These rings, when present in the area of interest in the reconstructed image, restrict the image analysis and limit the image processing efforts. This RAF is purely due to instrumental malfunctioning. The RAFs are a kind of high-frequency artifacts in the image, hence their removal is a paramount requirement. The RAF correction approaches a wide range of strategies \cite{pmb1}. 

Traditionally the RAF is removed via flat-field correction methodology \cite{r18}. This method involves measuring an image without subjecting a sample to the X-ray beam. The effects of all sorts of non-uniformities, like, the scintillator’s CCD detector’s non-uniform response, are recorded and considered as flat fields.  This flat field is then compared with the images received, and the necessary RAF is removed. The flat field correction is not very effective for RAF removal due to a wide range of detector response functions.
The hardware-based techniques \cite{r19, r20} invloves a nonstationary detector system that compensates for the malfunctioning of individual detector pixels. This results in an average detector performance, and a considerable reduction in RAF is reported. However, specialized hardware is needed to carry out this kind of RAF correction.  The third category includes image-based processing techniques. These are further classified into tomogram-based (post-processing of CT data) techniques \cite{jnst1,r36,r37,r38,r39,r40} and sinogram-based (pre-processing) techniques \cite{r15,r21}. Sinogram-based methods work directly with the sinogram data and aim to filter out the artifacts using low pass filters \cite{r22,r23,r24,r25}. The ring artifacts appear as straight lines in a vertical direction on the sinogram, making their recognition and interpretation easier when using sinogram-based algorithms. Furthermore, iterative systems based on relative total variations (RTV) have been presented in \cite{r15}, where the intensity deviations smoothing approach and image inpainting are used in the correction process. Most of these, however, fail to remove the powerful artifacts related to dead detector elements or damaged areas on the scintillator, and the majority of these techniques are only effective against specific kinds of stripes. The limitation of the discussed methods gives a strong motivation for the development of the deep-learning (DL) based ring artifacts removal procedure. Unlike filtering and hardware-based techniques, Machine Learning techniques have proven very promising in reducing these artifacts because of their capacity for feature detection and extraction, robustness to noise, flexibility, speed and efficiency, generalization, and other factors.

The DL-based techniques have shown promising results in artifact removal for other applications like image \cite{r26, r27, r28, r29, r30}. A correction method based on a residual neural network is proposed in \cite{r28}, where the artifacts correction network uses complementary information of each wavelet coefficient and a residual mechanism of the residual block to obtain high-precision artifacts through low operation costs. In \cite{r29}, a general open framework for MAR, which adopts the CNN to distinguish tissue structures from artifacts, is discussed. 
The approach involves two phases: CNN training and MAR. In the CNN training phase, a database with various CT images is built, and image patches are used to train a CNN. In the MAR phase, the trained CNN is employed to generate corrected images with reduced artifacts, followed by additional artifact reduction through replacing metal-affected projections and FBP reconstruction. CNN has been used in \cite{r30} to refine the performance of the normalized metal artifacts reduction (NMAR) method. A CNN-Based Hybrid Ring artifact Reduction Algorithm for CT Images has been proposed in \cite{r26} which uses image mutual correlation to generate a hybrid corrected image by fusing the information from ring artifacts reduction in the sinogram domain and output given by CNN. Polar coordinate transformation using a radial basis function neural network (RBFNN) to remove ring artifacts is proposed in \cite{r27}. Ring artifacts are transformed into linear artifacts by polar coordinate transformation and smoothing operators are applied to locate them exactly. Subsequently, RBFNN was operated on each linear artifact.

The biggest problem with the SXMCT images is that these images contain different backgrounds (attributed to different samples) and several artifacts, including the Ring artifacts. Therefore the individual study for the ring RAF removal is not possible. The article puts forward a DL-based ring artifacts removal approach which is relatively faster and has a wide range of applicability. The article also addresses the training data issue. The design procedure of purpose-specific synthetic data, hand-crafted data of SXMCT having only ring artifacts is described. The procedure assists in generating the required large diversity (wide diversity in terms of brightness, shape, size, and location) and volume hand-crafted SXMCT data with experimental inputs. The results of the DL-based approach have also been compared with the other conventional non-ML techniques to check the efficiency of our models. The detailed methodology, including the synthetic data generation along with the proposed Deep-Learning Architecture, is presented in section II. Section III contains the results and discussion, followed by conclusions in section IV.

\begin{figure*}[htbp]
  \centering
  \includegraphics[scale=0.35]{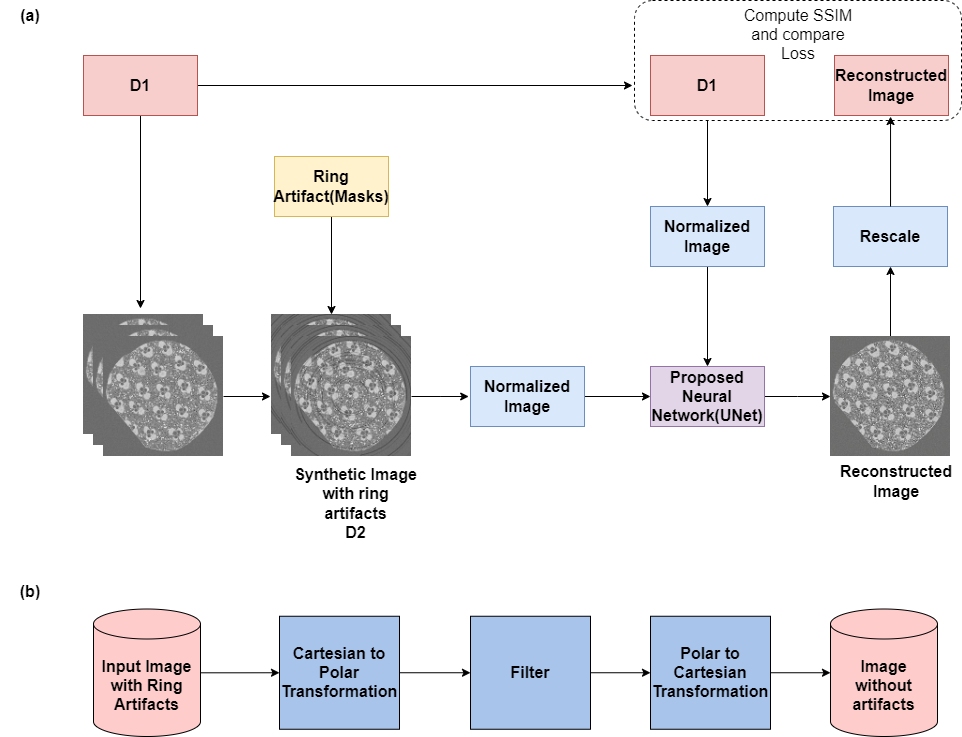}
  \caption{(a) Flowchart for ring artifact simulation and its removal, where D1 is the set of micro-ct images without ring artifacts, and D2 is the set of synthetic images having ring artifacts.
  (b) Flowchart of Non-ML method.}
  \label{fig:flowchart}
\end{figure*}

\section{METHODOLOGY}
Figure 1 represents the actual sample of the SXMCT image containing RAF. As visible in the figure, ring thickness, the distance between the rings, ring color, azimuthal angle of rings, and total number of rings are the important features.
The complete process followed in RAF removal consists of the steps shown in Figure 2. We have explored two methodologies, DL-based technique and non-ML techniques. As seen in Figure 2(a), in the DL-based approach, we begin with synthetic data generation, and then pre-processing of the synthetic images is done where they are normalized. Further, the network is trained on these images, and finally, the network is evaluated based on several parameters. In our work, we have used the UNet model for RAF correction. The key idea behind UNet is to use a Convolutional Neural Network (CNN) with both an encoding and a decoding path, with skip connections that allow the network to propagate high-resolution features from the encoder to the decoder. The UNet architecture can be used to reconstruct high-quality images from an incomplete or noisy input image, which is often achieved by training the UNet on a set of paired images, where the input images consist of degraded versions of the corresponding ground truth images as well as the ground truth images. To learn a mapping from the degraded input images to the high-quality ground truth images is the main goal. During training, the UNet learns to identify the relevant features in the input image and uses them to reconstruct the missing or degraded parts of the picture. The skip connections in the UNet architecture allow the network to preserve the high-frequency details of the image, which is essential for reconstructing sharp and accurate images. During training, UNet is typically optimized using a loss function such as MSE, which compares the network's output to the ground truth image and penalizes their differences. The network weights are updated using backpropagation, which computes the gradient of the loss concerning the network parameters and uses it to update the consequences in the direction that minimizes the loss. Once trained,  it can reconstruct new images by applying the learned mapping to new degraded input images.\\
In non-ML approach, as seen in Figure 2(b), the image with ring artifacts is converted from cartesian to polar coordinate system. In Cartesian coordinates, the image is represented in terms of x and y coordinates, forming a grid-like structure. In polar coordinates, the representation is based on the radial distance from a central point and the angle at which a point is located relative to that central point. Further, the filter is applied to this image. Again, the filtered image is transformed from polar coordinates to cartesian coordinates.\\
The performance of both approaches has been evaluated using SSIM and MSE. The dataset used in the research work consists of the images obtained by performing micro-CT of the ice cream stick made up of bamboo. By infusing the domain knowledge, we have synthetically produced the images with the ring artifacts using the above-mentioned features.

\begin{figure}[!htbp]
    \centering
    \includegraphics[scale=0.45]{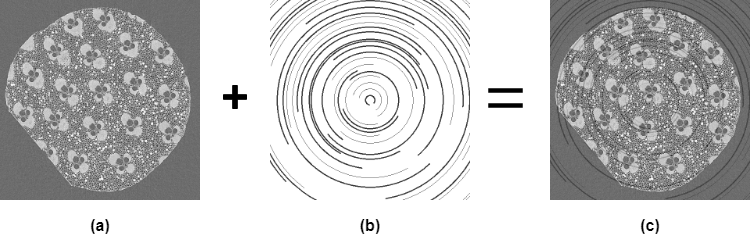}
    \caption{Synthetic Data Generation Process}
    \label{fig:process}
\end{figure}
\vspace{-10pt}
\subsection{Synthetic data generation}

Figure 3 shows the high-level process followed in synthetic data generation. We have an experimental image from the setup that can generate micro-ct images without RAF. This image is superimposed with synthetic ring artifact masks, thereby producing the synthetic image with RAF.

For a similar simulation of ring artifacts in the sample images, we create concentric rings on a white background using the below five features of the ring.

\begin{enumerate}[label=\roman*)]
\item Color of the rings(grayscale): It defines the color of the concentric circles generated; the closer it is to 255, the lighter will be the color of the rings.

\item Distance between two concentric rings: This can be done by random initialization of the radius of concentric circles

\item Azimuthal angle of the rings: It defines the start angle and end angle of the ring

\item Total number of rings in the image: This parameter defines the number of concentric circles we want to generate in the image

\item Ring thickness: By randomly varying the thickness of the rings, new rings can be generated within the given range of thickness 
\end{enumerate}
\vspace{-3pt}

\begin{figure}[!htbp]
\centering
        \includegraphics[scale=0.4]{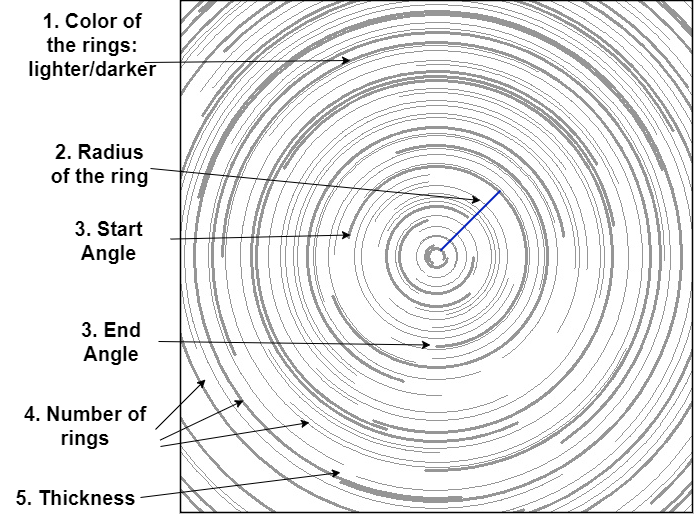}
        \caption{Important features for ring artifacts creation}
        \label{fig:parameters}
\end{figure}


\noindent Figure 4 shows the parameters used for creating the concentric rings. By randomly initializing these parameters, we have created 25 such masks, which are then used to generate the synthetic training data.

As discussed in the above section, circular ring artifacts (i.e., masks) have been created by varying the 5 features. The experimental images have been obtained by performing micro-CT of bamboo ice-cream stick. We have superimposed these images with the masks which are generated synthetically by the above methods. We have superimposed 25 such masks generated by varying the above parameters with 101 sample images provided. Thus, 2525 diverse images are developed because in order to ensure that the trained network can handle images other than the original training dataset, it is necessary to expose it to a diverse dataset while training. Figure 5 shows a few synthetic data samples. The training dataset consists of 2525 pairs of images where one is with RAF and the other is without RAF which were directly obtained from the experiment.

\begin{figure}[!htbp]
    \centering
    \includegraphics[scale=0.5]{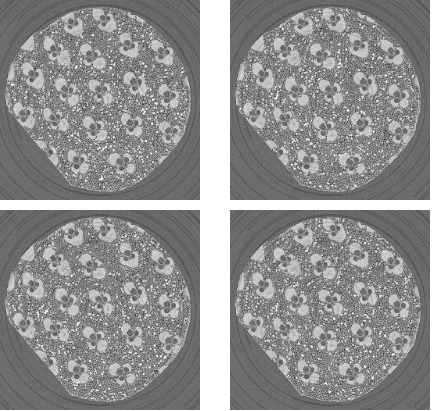} 
    \caption{Samples of synthetic images with ring artifacts}
    \label{fig:train_samples}
\end{figure}

\begin{figure*}[ht]
  \centering
  \includegraphics[scale=0.37]{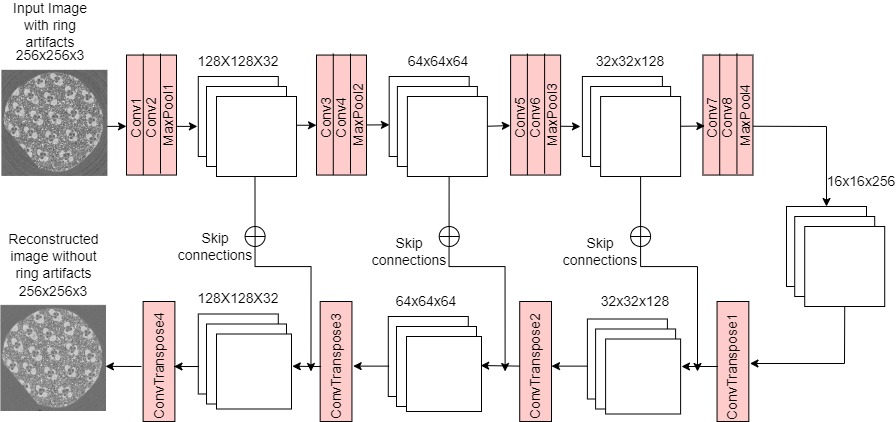}
  \caption{Proposed encoder-decoder based architecture with skip connections.}
  \label{fig:unet_arch}
\end{figure*}

\subsection{Deep-Learning Architecture}


This section proposes a DL-based architecture for denoising of micro-ct images. The proposed architecture is CNN-based UNet, where the input to the network is 256x256 RGB normalized image, which is synthetically generated by adding ring artifacts as noise to the SXMCT images provided. The encoder block consists of a convolution block containing a convolution layer followed by ReLU(Rectified Linear Unit) activation and batch normalization followed by the same three layers mentioned above. Further, max-pooling of kernel size 2x2 is carried out, which is used to downsample the images. There are N such encoder units. Each decoder unit consists of a transposed convolution with a kernel size of 2x2. The input to the decoder layer is directly connected to the output of the encoder layer using skip connections. Skip connections are implemented by concatenating the result of one layer to the other layer to which it is connected. The concatenation is followed by a dropout layer with a rate of 0.3 and a convolution block.

The normalized image X is given as input to the network, and the network learns its parameters by minimizing the loss function between the actual image without ring artifacts, denoted as Y, and the output of the reconstructed image without ring artifacts, denoted as F(X;$\theta$). The loss function for training the architecture is given as follows:
\begin{equation}
L(\theta)  = \frac{1}{N} \sum_{i=1}^{N} \|F(X;\theta)- Y\|^2_2 + \lambda (1 - p)^2 \sum_{j=1}^{l} \|w_j\|^2_2
\end{equation}

where N = the number of training samples, $\lambda$ is the regularization strength, p is the probability that a neuron will be dropped out, and w is the weight of a neuron

\begin{equation}
L(\theta)  = \frac{1}{N} \sum_{i=1}^{N} (y_i - \hat{y}_i)^2 + \lambda (1 - p)^2 \sum_{j=1}^{M} (w_j^2)
\end{equation}

Figure 6 represents the detailed architecture of the proposed model.

\subsection{Non-ML methods}
The non-ML approach includes the application of various image processing filters. A filter is a mathematical operation applied to a set of data, such as an image or a signal, to enhance or suppress certain features. Filters are used to modify the information content of the data, and they can be categorized into various types based on their functions. 
Each category of filter is designed to address specific aspects of image processing, noise reduction, edge enhancement, or feature extraction. The choice of filter depends on the characteristics of the data and the goals of the image processing task. Here, we have used the Fast Fourier Transform (FFT) filter, Bilateral filter, Stripe Filter, and Butterworth filter and compared the results with the results of DL-based approaches. FFT-based methods rely on the assumption that ring artifacts correspond to high-frequency components in the Fourier domain. As a result, they can be removed by damping these components\cite{r31}. Butterworth filter aids in suppression by selectively filtering specific frequencies associated with ring artifacts, leading to improved image quality and is characterized by its ability to offer a customizable frequency response\cite{r32}. Bilateral filtering uses a nonlinear combination of adjacent image values to smooth images while maintaining edges\cite{r33,r34}. Stripe filter can be used to eliminate stripe artifacts and ring artifacts and it is based on wavelet decomposition and Fourier filtering\cite{r35}.

\subsection{Performance Metrics}

Performance evaluation of the DL-based technique and non-ML techniques has been done using SSIM and MSE is used as a loss function.
SSIM is a widely used metric to evaluate the similarity between two images.

\begin{equation}
\text{SSIM}(x, y) = \frac{{\left(2\mu_x\mu_y + C_{1}\right)\left(2\sigma_{xy} + C_{2}\right)}}{{\left(\mu_x^2 + \mu_y^2 + C_1\right)\left(\sigma_x^2 + \sigma_y^2 + C_2\right)}}
\end{equation}

where $\mu_{x}$, $\mu_{y}$ are the mean values of the pixels in images x and y, respectively. $\sigma^{2}_{x}$, $\sigma^{2}_{y}$  are the variances of the pixels in images x and y, respectively. $\sigma_{xy}$ is the covariance between the pixels in images x and y, and C1 and C2 are constants added to avoid division by zero.
The numerator of the equation measures the similarity in terms of luminance, contrast, and structure between the two images. At the same time, the denominator normalizes the similarity measure to ensure that the values are between -1 and 1.
SSIM values range between -1 and 1, where a value of 1 indicates that the two images are identical. A value of 0 indicates that the two images are entirely dissimilar, and a value of -1 indicates perfect anti-correlation. 

MSE is a commonly used loss function in image reconstruction tasks because it measures the average squared difference between predicted and actual pixel values. MSE penalizes significant errors more heavily than minor errors, and the goal of image reconstruction is typically to minimize the overall difference between the predicted and actual images. A higher value of MSE designates a more significant difference between the original image and the processed image. The formula is as follows :
\begin{equation}
\text{MSE} = \frac{1}{N^2}\sum_{i=1}^{N}\sum_{j=1}^{N}\left(B_{ij} - A_{ij}\right)^2 
\end{equation}
$B_{ij}$ are the predicted pixel values at i and j indices and $A_{ij}$ are the actual pixel values at i and j indices, and n is the total number of pixels in the image.
\\
\vspace{-3pt}
\section{Experimental Results and Discussions}

In this section, we discuss the training details and computational experiments, performance evaluation of the results obtained in the experiment, testing details, feature map visualization, and the ablation study carried out on the proposed network.

\subsection{DL-based RAF correction method}
Initially, the experiment was carried out with 4 encoder-decoder architecture as shown in Figure 6 followed by the experiments with 5 encoder-decoder and 6 encoder-decoder architecture. However, the best results are obtained for the architecture with 6 encoder-decoder units. The model is trained for 100 epochs on GPU P100 using Keras API with TensorFlow running in the backend. Adam optimizer has been used for training with a learning rate of 0.001, and $\varepsilon$ is $10^{-7}$. The training SSIM obtained is 0.9440, and the MSE is 0.0013.
Figure 7 represents the results obtained using the trained network.

\begin{figure}[!htbp]
    \centering
    \includegraphics[scale=0.5]{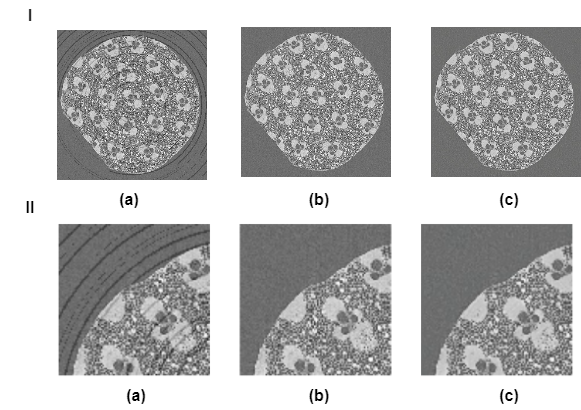}
    \caption{Comparison of  (I) (a)Input, (b)Actual, and (c)Output images and (II) 100x100 sized patch of (a)Input, (b)Actual, and (c)Output images obtained through the trained network}
    \label{fig:phase2_patch}
\end{figure}

We have generated different datasets synthetically for testing purposes. For that, we have varied one parameter at a time, and the rest of the features have been kept static. However, the concentric ring radius has been varied for all the datasets. Firstly, we varied the radii of rings and created a testing dataset of 303 images. Similarly, the second dataset has been generated by randomizing the thickness and radius alone. Finally, we have generated the third testing dataset by varying the number of rings and their radii.
The results obtained for each of the datasets mentioned above are shown in Table 1.

\renewcommand{\arraystretch}{2.5}
\begin{table}[!htbp]
\centering
\caption{Results of various testing datasets}
\begin{tabular}{|c|c|c|}
\hline
Varying parameter & Test SSIM & Test MSE  \\ \hline
Radius & 0.9173 &  0.0021 \\ \hline
Radius and thickness & 0.9083 & 0.0023  \\ \hline
Radius and No. of rings & 0.94647 & 0.00080  \\ \hline
\end{tabular}
\label{test}
\end{table}
\smallskip\smallskip\smallskip

\begin{figure*}[!htbp]
    \centering
    \includegraphics[scale=0.3]{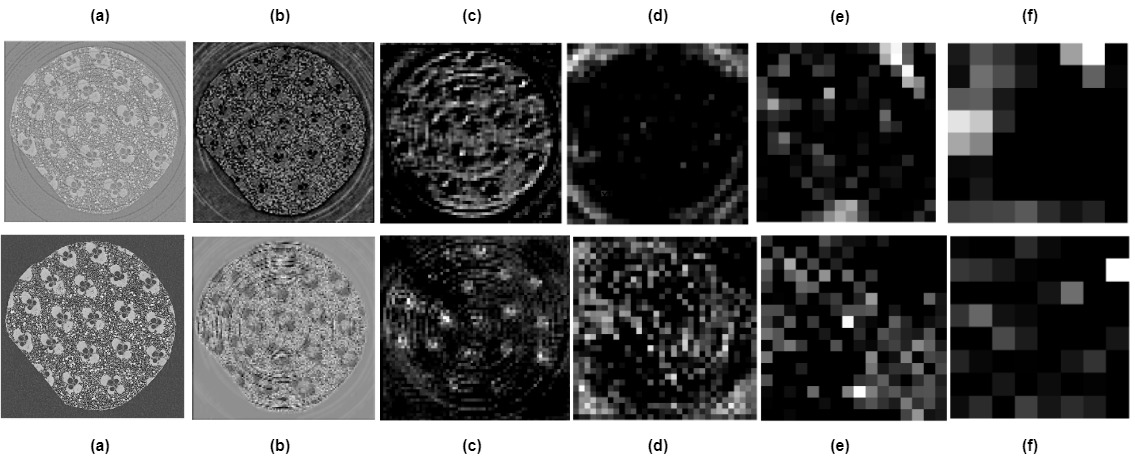}
    \caption{Row-1 represents feature maps of the output of the six encoder units; Row-2 represents feature maps of the output of the six encoder units; The corresponding (a-f) images represent the encoder-decoder pairs for the feature output in spatial dimension}
    \label{fig:feature_map}
\end{figure*}

Figure 8 is the network visualization of the feature map, which helps to understand the output after each encoder-decoder unit. The encoder downsamples the image from (a) to (f), whereas the decoder upsamples it from (f) to (a). The encoder learns the semantic and structural features of the object, whereas the decoder learns the pattern for the removal of ring artifacts. The total number of feature maps for a layer equals the number of filters used in the convolution or transposed convolution layer. It can be observed that the encoder initially learns high-level features followed by low-level features, whereas the decoder initially learns low-level features followed by high-level features. 
Initial encoders retain most of the input image features. This gives us an intuition that these initial filters might be primitive edge detectors. 
As we go deeper, the features extracted by the filters become visually less interpretable due to the pooling operations. The pooling operations reduce the resolution of the feature maps while increasing the neurons' receptive field. Consequently, the information in the deeper layers becomes more abstract and less spatially localized. As we move on to the decoders, the initial decoder layers receive the most abstract and downsampled feature maps from the corresponding encoder layers and hence the feature maps produced are less interpretable. As we move forward, the upsampling techniques increase the spatial resolution of the feature maps. It also concatenates them with the corresponding feature maps from the encoder through skip connections. This merging of feature maps helps to reintroduce spatial information and low-level details that are crucial for accurate reconstruction.

\subsection{Ablation study}
In order to explore the effect of changing the model parameters like number of encoder and decoder blocks, upsampling methods on the predicted SSIM and MSE losses of the reconstructed images an ablation study has been carried out on the generated dataset.

\subsubsection{Effect of encoder and decoder blocks}
In this study, we have performed additional experiments with four, five, and six encoder-decoder units. For six encoders, we have obtained an SSIM value of 0.9523 and an MSE loss value of 0.0011, which is a good indicator for image reconstruction by the network. By comparing with the other models, in terms of SSIM and loss, as shown in Table 2, we found that the model's performance with six encoders was the best, and hence the proposed architecture of the network contains six encoder-decoder. 

\begin{table}[!htbp]
\centering
\caption{Results of varying the number of Encoder-Decoder units}
\small
\begin{tabular}{|c|c|c|}
\hline
Encoder-Decoder units & Validation SSIM & Validation Loss \\ \hline
6 units & 0.9523 & 0.0011 \\ \hline
5 units & 0.9611 & 0.0013 \\ \hline
4 units & 0.9545 & 0.0033 \\ \hline
\end{tabular}
\label{AS1}
\end{table}

\subsubsection{Effect of Up-sampling method}
Here, we have systematically investigated the impact of employing various upsampling methods. The upsampling layers follow an interpolation scheme, which increases the input dimension. Various upsampling methods like transposed convolution, bicubic, nearest, and 
bilinear interpolation have been used to analyze their effect on model performance. In Table 3, we have presented the outcomes for all the upsampling methods. It is evident from the results that the transposed convolution yields the most favorable outcomes, and hence, it has been used in the proposed architecture of the network.

\begin{table}[!htbp]
\centering
\caption{Impact of using various Up-Sampling methods}
\begin{tabular}{|c|c|c|c|c|}
\hline
Upsampling Methods & Validation SSIM & Validation Loss \\ \hline
Transposed Conv & 0.9523 & 0.0011 \\ \hline
Bilinear Interpolation & 0.9105 & 0.0049 \\ \hline
Bicubic Interpolation & 0.9451 & 0.0023 \\ \hline
Nearest Interpolation & 0.9143 & 0.0079 \\ \hline
\end{tabular}
\label{AS2}
\end{table}

\subsection{Comparison with Non-ML methods}

Various filters Butterworth filter, Fast-Fourier Transform filter, Bilateral filter, and Stripe filter, have been applied to the synthetic images and figure 9 shows the reconstructed image obtained from the network as well as by the application of these filters and their corresponding SSIM value. As it can be observed, among various non-ML methods applied, Stripe filter works better than the other filters without significantly degrading the image quality or blurring important details. However, it is evident that UNet outperforms all the other conventional techniques by giving a comparatively cleaner image and the highest SSIM value.

 \begin{figure*}[!htbp]
    \centering
    \includegraphics[width=4.5in,height=1.14in]
    {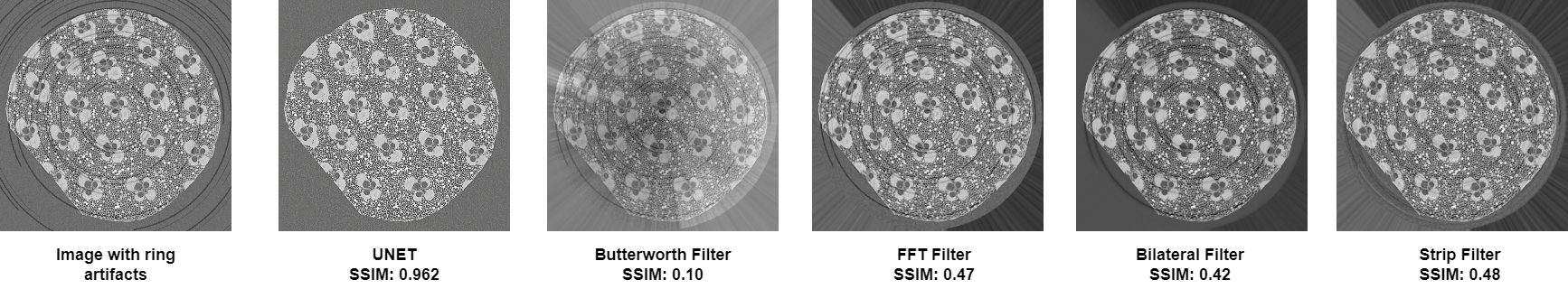}
    \caption{Comparison of ML (UNET) vs various non-ML (filter based) techniques}
    \label{fig:non_ml_result}
\end{figure*}

\section{Conclusion}

This study introduces a CNN-based deep learning model inspired by UNet with a series of encoder-decoder units with skip connections to denoise the micro-CT images with a specific goal of removing ring artifacts. Proposed approach involves training the model with a dataset consisting of synthetic images which is generated by masking the micro-ct images (without ring artifact) with concentric ring structures. These ring structures are simulated by varying parameters like ring thickness, the distance between two concentric rings, azimuthal angle, the total number of rings in the image, and the color of the rings. The trained network is then used to reconstruct the synthetic images which removes the ring artifacts. The results have also been compared with various non-ML techniques. The results of both approaches have been evaluated using various metrics such as SSIM and MSE. For the DL based techniques, the average SSIM performance metrics is about 0.9523, and the average MSE is about 0.0011. Testing carried out on different datasets has been discussed. Ablation studies on the effect of encoder-decoder units and other upsampling methods have also been discussed. To get a detailed idea of how the network is removing the RAF, feature map visualization has also been performed. We observe that, DL based technique offer definite advantages over non-ML approaches. Because of end-to-end learning capabilities of DL model, the complex and varied nature of ring artifacts can be captured effectively without the need for explicit filtering methods or manual feature design. Moreover, our results show that machine learning based approach do exceptionally well in feature learning, automatically obtaining hierarchical features that are essential for handling the variety of attributes of ring artifacts. They can generalize well to the unseen pattern. Since DL model can handle complexity and non-linear relationships, they perform better than traditional techniques, which could find it difficult to deal with the complex nature of ring artifacts.
In contrast to conventional methods for addressing ring artifacts, the methodology presented in this study offers a more efficient solution in removing such artifacts. The findings underscore that utilizing a deep learning-based approach not only offers a more efficient solution but also serves as a viable alternative to traditional non-machine learning methods.

\bibliographystyle{unsrt}

\nocite{*}
\bibliography{article}

\begin{thebibliography}{10}

\bibitem{r1}
Philip~J Withers, Charles Bouman, Simone Carmignato, Veerle Cnudde, David Grimaldi, Charlotte~K Hagen, Eric Maire, Marena Manley, Anton Du~Plessis, and Stuart~R Stock.
\newblock X-ray computed tomography.
\newblock {\em Nature Reviews Methods Primers}, 1(1):18, 2021.

\bibitem{r2}
Douglas~P Boyd.
\newblock Computed tomography: physics and instrumentation.
\newblock {\em Academic Radiology}, 2:S138--S140, 1995.

\bibitem{r3}
Lee~W Goldman.
\newblock Principles of ct and ct technology.
\newblock {\em Journal of nuclear medicine technology}, 35(3):115--128, 2007.

\bibitem{r4}
Karen~S Caldemeyer and Kenneth~A Buckwalter.
\newblock The basic principles of computed tomography and magnetic resonance imaging.
\newblock {\em Journal of the American Academy of Dermatology}, 41(5):768--771, 1999.

\bibitem{r5}
Christy~A Hipsley, Rocio Aguilar, Jay~R Black, and Scott~A Hocknull.
\newblock High-throughput microct scanning of small specimens: preparation, packing, parameters and post-processing.
\newblock {\em Scientific reports}, 10(1):13863, 2020.

\bibitem{r6}
Kleoniki Keklikoglou, Christos Arvanitidis, Georgios Chatzigeorgiou, Eva Chatzinikolaou, Efstratios Karagiannidis, Triantafyllia Koletsa, Antonios Magoulas, Konstantinos Makris, George Mavrothalassitis, Eleni-Dimitra Papanagnou, et~al.
\newblock Micro-ct for biological and biomedical studies: A comparison of imaging techniques.
\newblock {\em Journal of imaging}, 7(9):172, 2021.

\bibitem{r7}
Aymeric Larrue, Aline Rattner, Zsolt-Andrei Peter, C{\'e}cile Olivier, Norbert Laroche, Laurence Vico, and Fran{\c{c}}oise Peyrin.
\newblock Synchrotron radiation micro-ct at the micrometer scale for the analysis of the three-dimensional morphology of microcracks in human trabecular bone.
\newblock {\em PLoS one}, 6(7):e21297, 2011.

\bibitem{r8}
Charlotta~K{\"a}mpfe Nordstr{\"o}m, Hao Li, Hanif~M Ladak, Sumit Agrawal, and Helge Rask-Andersen.
\newblock A micro-ct and synchrotron imaging study of the human endolymphatic duct with special reference to endolymph outflow and meniere’s disease.
\newblock {\em Scientific Reports}, 10(1):8295, 2020.

\bibitem{r9}
Camilla~Albeck Neldam, Torsten Lauridsen, Alexander Rack, Tore~Tranberg Lefolii, Niklas~Rye J{\o}rgensen, Robert Feidenhans, and Else~Marie Pinholt.
\newblock Application of high resolution synchrotron micro-ct radiation in dental implant osseointegration.
\newblock {\em Journal of Cranio-Maxillofacial Surgery}, 43(5):682--687, 2015.

\bibitem{r10}
Christian Norvik, Christian~Karl West{\"o}{\"o}, Niccol{\`o} Peruzzi, Goran Lovric, Oscar van~der Have, Rajmund Mokso, Ida Jeremiasen, Hans Brunnstr{\"o}m, Csaba Galambos, Martin Bech, et~al.
\newblock Synchrotron-based phase-contrast micro-ct as a tool for understanding pulmonary vascular pathobiology and the 3-d microanatomy of alveolar capillary dysplasia.
\newblock {\em American Journal of Physiology-Lung Cellular and Molecular Physiology}, 318(1):L65--L75, 2020.

\bibitem{r11}
Kaan Orhan, Karla de~Faria~Vasconcelos, and Hugo Ga{\^e}ta-Araujo.
\newblock Artifacts in micro-ct.
\newblock {\em Micro-computed Tomography (micro-CT) in Medicine and Engineering}, pages 35--48, 2020.

\bibitem{r12}
Mohamed~Elsayed Eldib, Mohamed Hegazy, Yang~Ji Mun, Myung~Hye Cho, Min~Hyoung Cho, and Soo~Yeol Lee.
\newblock A ring artifact correction method: Validation by micro-ct imaging with flat-panel detectors and a 2d photon-counting detector.
\newblock {\em Sensors}, 17(2):269, 2017.

\bibitem{r13}
Anton Du~Plessis, Chris Broeckhoven, Anina Guelpa, and Stephan~Gerhard Le~Roux.
\newblock Laboratory x-ray micro-computed tomography: a user guideline for biological samples.
\newblock {\em Gigascience}, 6(6):gix027, 2017.

\bibitem{r14}
F~Edward Boas, Dominik Fleischmann, et~al.
\newblock Ct artifacts: causes and reduction techniques.
\newblock {\em Imaging Med}, 4(2):229--240, 2012.

\bibitem{r15}
Jakub {\v{S}}alplachta, Tom{\'a}{\v{s}} Zikmund, Marek Zemek, Adam B{\v{r}}{\'\i}nek, Yoshihiro Takeda, Kazuhiko Omote, and Jozef Kaiser.
\newblock Complete ring artifacts reduction procedure for lab-based x-ray nano ct systems.
\newblock {\em Sensors}, 21(1):238, 2021.

\bibitem{jnst1}
Zhouping Wei, Sheldon Wiebe, and Dean Chapman.
\newblock Ring artifacts removal from synchrotron ct image slices.
\newblock {\em Journal of Instrumentation}, 8:C06006, 06 2013.

\bibitem{r16}
Nghia~T Vo, Robert~C Atwood, and Michael Drakopoulos.
\newblock Superior techniques for eliminating ring artifacts in x-ray micro-tomography.
\newblock {\em Optics express}, 26(22):28396--28412, 2018.

\bibitem{r17}
Emran Mohammad~Abu Anas, Soo~Yeol Lee, and Md~Kamrul Hasan.
\newblock Classification of ring artifacts for their effective removal using type adaptive correction schemes.
\newblock {\em Computers in biology and medicine}, 41(6):390--401, 2011.

\bibitem{pmb1}
Emran Anas, Soo~Yeol Lee, and Md~Kamrul Hasan.
\newblock Removal of ring artifacts in ct imaging through detection and correction of stripes in the sinogram.
\newblock {\em Physics in medicine and biology}, 55:6911--30, 11 2010.

\bibitem{r18}
Matthias Ruf and Holger Steeb.
\newblock An open, modular, and flexible micro x-ray computed tomography system for research.
\newblock {\em Review of Scientific Instruments}, 91(11), 2020.

\bibitem{r19}
Yining Zhu, Mengliu Zhao, Hongwei Li, and Peng Zhang.
\newblock Micro-ct artifacts reduction based on detector random shifting and fast data inpainting.
\newblock {\em Medical physics}, 40(3):031114, 2013.

\bibitem{r20}
GR~Davis and JC~Elliott.
\newblock X-ray microtomography scanner using time-delay integration for elimination of ring artefacts in the reconstructed image.
\newblock {\em Nuclear Instruments and Methods in Physics Research Section A: Accelerators, Spectrometers, Detectors and Associated Equipment}, 394(1-2):157--162, 1997.

\bibitem{r36}
Jan Sijbers and Andrei Postnov.
\newblock Reduction of ring artefacts in high resolution micro-ct reconstructions.
\newblock {\em Physics in Medicine \& Biology}, 49(14):N247, 2004.

\bibitem{r37}
Yiannis Kyriakou, Daniel Prell, and Willi~A Kalender.
\newblock Ring artifact correction for high-resolution micro ct.
\newblock {\em Physics in medicine \& biology}, 54(17):N385, 2009.

\bibitem{r38}
Xiaokun Liang, Zhicheng Zhang, Tianye Niu, Shaode Yu, Shibin Wu, Zhicheng Li, Huailing Zhang, and Yaoqin Xie.
\newblock Iterative image-domain ring artifact removal in cone-beam ct.
\newblock {\em Physics in Medicine \& Biology}, 62(13):5276, 2017.

\bibitem{r39}
Luxin Yan, Tao Wu, Sheng Zhong, and Qiude Zhang.
\newblock A variation-based ring artifact correction method with sparse constraint for flat-detector ct.
\newblock {\em Physics in Medicine \& Biology}, 61(3):1278, 2016.

\bibitem{r40}
ANM Ashrafuzzaman, Soo~Yeol Lee, and Md~Kamrul Hasan.
\newblock A self-adaptive approach for the detection and correction of stripes in the sinogram: suppression of ring artifacts in ct imaging.
\newblock {\em EURASIP Journal on Advances in Signal Processing}, 2011:1--13, 2011.

\bibitem{r21}
Dong-Jiang Ji, Gang-Rong Qu, Chun-Hong Hu, Bao-Dong Liu, Jian-Bo Jian, and Xiao-Kun Guo.
\newblock Anisotropic total variation minimization approach in in-line phase-contrast tomography and its application to correction of ring artifacts.
\newblock {\em Chinese Physics B}, 26(6):060701, 2017.

\bibitem{r22}
Carsten Raven.
\newblock Numerical removal of ring artifacts in microtomography.
\newblock {\em Review of scientific instruments}, 69(8):2978--2980, 1998.

\bibitem{r23}
Beat M{\"u}nch, Pavel Trtik, Federica Marone, and Marco Stampanoni.
\newblock Stripe and ring artifact removal with combined wavelet—fourier filtering.
\newblock {\em Optics express}, 17(10):8567--8591, 2009.

\bibitem{r24}
Fazle Sadi, Soo~Yeol Lee, and Md~Kamrul Hasan.
\newblock Removal of ring artifacts in computed tomographic imaging using iterative center weighted median filter.
\newblock {\em Computers in biology and medicine}, 40(1):109--118, 2010.

\bibitem{r25}
Mohamed~Elsayed Eldib, Mohamed Hegazy, Yang~Ji Mun, Myung~Hye Cho, Min~Hyoung Cho, and Soo~Yeol Lee.
\newblock A ring artifact correction method: Validation by micro-ct imaging with flat-panel detectors and a 2d photon-counting detector.
\newblock {\em Sensors}, 17(2):269, 2017.

\bibitem{r26}
Shaojie Chang, Xi~Chen, Jiayu Duan, and Xuanqin Mou.
\newblock A cnn-based hybrid ring artifact reduction algorithm for ct images.
\newblock {\em IEEE Transactions on Radiation and Plasma Medical Sciences}, 5(2):253--260, 2021.

\bibitem{r27}
Zhen Chao and Hee-Joung Kim.
\newblock Removal of computed tomography ring artifacts via radial basis function artificial neural networks.
\newblock {\em Physics in Medicine \& Biology}, 64(23):235015, 2019.

\bibitem{r28}
Tianyu Fu, Yan Wang, Kai Zhang, Jin Zhang, Shanfeng Wang, Wanxia Huang, C~Yao, C~Zhou, and Q~Yuan.
\newblock Deep-learning-based ring artifact correction for tomographic reconstruction.
\newblock {\em Journal of Synchrotron Radiation}, 30(3), 2023.

\bibitem{r29}
Yanbo Zhang and Hengyong Yu.
\newblock Convolutional neural network based metal artifact reduction in x-ray computed tomography.
\newblock {\em IEEE transactions on medical imaging}, 37(6):1370--1381, 2018.

\bibitem{r30}
Lars Gjesteby, Qingsong Yang, Yan Xi, Hongming Shan, Bernhard Claus, Yannan Jin, Bruno De~Man, and Ge~Wang.
\newblock Deep learning methods for ct image-domain metal artifact reduction.
\newblock In {\em Developments in X-ray Tomography XI}, volume 10391, pages 147--152. SPIE, 2017.

\bibitem{r31}
Nghia~T. Vo, Robert~C. Atwood, and Michael Drakopoulos.
\newblock Superior techniques for eliminating ring artifacts in x-ray micro-tomography.
\newblock {\em Opt. Express}, 26(22):28396--28412, Oct 2018.

\bibitem{r32}
Li~Zhongshen.
\newblock Design and analysis of improved butterworth low pass filter.
\newblock In {\em 2007 8th International Conference on Electronic Measurement and Instruments}, pages 1--729--1--732, 2007.

\bibitem{r33}
C.~Tomasi and R.~Manduchi.
\newblock Bilateral filtering for gray and color images.
\newblock In {\em Sixth International Conference on Computer Vision (IEEE Cat. No.98CH36271)}, pages 839--846, 1998.

\bibitem{r34}
Juan C.~Ramirez Giraldo, Zachary~S. Kelm, Luis~S. Guimaraes, Lifeng Yu, Joel~G. Fletcher, Bradley~J. Erickson, and Cynthia~H. McCollough.
\newblock Comparative study of two image space noise reduction methods for computed tomography: Bilateral filter and nonlocal means.
\newblock In {\em 2009 Annual International Conference of the IEEE Engineering in Medicine and Biology Society}, pages 3529--3532, 2009.

\bibitem{r35}
Beat M{\"u}nch, Pavel Trtik, Federica Marone, and Marco Stampanoni.
\newblock Stripe and ring artifact removal with combined wavelet—fourier filtering.
\newblock {\em Optics express}, 17(10):8567--8591, 2009.

\end{thebibliography}

\end{document}